\documentclass[12pt,preprint]{aastex}
\usepackage{emulateapj5}
\usepackage{apjfonts}
\usepackage{graphicx}
\usepackage{amsfonts}

\slugcomment{ApJ Letters, scheduled for 2003 January 10th}

\def\r14{$r^{1/4}$}
\def\rn{$r^{1/n}$}
\def\kms{km\,s$^{-1}$}
\def\nhst{$n_{hst}$}
\def\ngb{$n_{gb}$}

\shorttitle{Bulge NIR nuclear profiles}
\shortauthors{Balcells, Graham, Dom\'\i nguez-Palmero \& Peletier}

\begin{document}

\doublespace 

%%%%%%%%%%%%%%%
% Title and authors
%%%%%%%%%%%%%%%

\title{Galactic bulges from {\it Hubble Space Telescope} NICMOS 
observations: the lack of $r^{1/4}$ bulges \altaffilmark{1}}

\author{Marc Balcells, Alister W. Graham\altaffilmark{2}, 
Lilian Dom\'\i nguez-Palmero}
\affil{Instituto de Astrof\'\i sica de Canarias, 38200 La Laguna, 
Tenerife, Spain}
\email{balcells@ll.iac.es}

\and

\author{Reynier F. Peletier\altaffilmark{3}}
\affil{School of Physics and Astronomy, University of Nottingham, NG7 2RD, UK}

\altaffiltext{1}{Based on observations made with the NASA/ESA {\sl 
Hubble Space Telescope}, obtained at the Space Telescope Science 
Institute, which is operated by the Association of Universities for 
Research in Astronomy, Inc., under NASA contract NAS 5-26555.}
\altaffiltext{2}{Present address: Department of Astronomy, 
University of Florida, Gainesville.}
\altaffiltext{3}{also: CRAL, Observatoire de Lyon, 9, 
Av. Charles Andr\'e, 69230 Saint Genis Laval, France}

\begin{abstract}
We use {\it Hubble Space Telescope} ({\it HST}) near-infrared imaging
to explore the shapes of the surface brightness profiles of bulges of
S0-Sbc galaxies at high resolution.  Modeling extends to the outer
bulge via bulge-disk decompositions of combined {\it HST} - ground
based profiles.  Compact, central unresolved components similar to
those reported by others are found in $\sim$84\% of the sample.  We
also detect a moderate frequency ($\sim$34\%) of nuclear components
with exponential profiles which may be disks or bars.  Adopting the
S\'ersic \rn\ functional form for the bulge, none of the bulges have
an \r14\ behaviour; derived S\'ersic shape-indices are $<n> = 1.7 \pm
0.7$.  For the same sample, fits to NIR ground-based profiles yield
S\'ersic indices up to $n = 4-6$.  The high-$n$ of ground-based
profiles are a result of nuclear point sources blending with the bulge
extended light due to seeing.  The low S\'ersic indices are not
expected from merger violent relaxation, and argue against significant
merger growth for most bulges.

\end{abstract}


\section{Introduction}
\label{Sec:Introduction}

Whether galaxy mergers (e.g.\ Kauffmann et al.\ 1996) or instead disk
instabilities (Pfenniger \& Norman 1990) may be viewed as the dominant
process of formation and growth of the central bulges in spiral and S0
galaxies depends to some extent on the shape of the surface brightness
profiles of bulges.  Simulations of mergers of similar-size galaxies
yield spheroidal systems following the \r14\ law (e.g.\ Barnes 1988). 
The accretion of dense satellites onto bulge-disk galaxies also leads
to \r14\ bulges (Aguerri, Balcells \& Peletier 2001, herafter ABP01). 
In contrast, disk instabilities lead to bulges with approximately
exponential profiles (Combes et al.\ 1990).

Over the past decade several studies have stressed that bulges of late-
to intermediate-type spiral galaxies are best fit with profiles of
exponential to $r^{1/2}$ shapes 
(Andredakis \& Sanders 1994; de Jong 1996; Courteau, de Jong \& Broeils 1996;
Seigar \& James 1998; Carollo, Stiavelli \& Mach 1998; MacArthur et al.\
2002).  However, many bulge-disk decompositions are still performed using \r14\
bulges only, especially at high-$z$, e.g. Simard et al.\ (2002); 
the \r14\ law is
often cited for massive or early-type bulges: Carollo and collaborators (e.g.\
Seigar et al.\ 2002 and references therein) routinely classify bulges into
\r14\ and exponential types. Do massive bulges follow the \r14\ model?  
Do bulges come in two flavors, perhaps pointing at massive bulges 
growing by early mergers and less-massive bulges growing by
disk instabilities?  

Fitting bulge profiles with the S\'ersic model $\mu(r) \sim r^{1/n}$ provides
an avenue for measuring the true shapes and concentrations of the light
distribution of bulges.  Various studies that include early-type bulges 
(Andredakis, Peletier \& Balcells
1995, hereafter APB95; Khosroshahi et al.\ 2000; Graham 2001; M\"ollenhoff \&
Heidt 2001) find a continuous distribution of S\'ersic shape indices $n$ that
scales with bulge luminosity from $n<1$ to $n>4$.  At the high-$n$ end, these
results depend strongly on the innermost regions of the profiles.  Inner disks,
compact sources and star formation, present in galaxy nuclei (e.g.\ Phillips et
al.\ 1996; Carollo et al.\ 2002; Rest et al.\ 2001; Pizzela et al.\ 2002), contribute
light that cannot be disentangled from the spheroid light at ground-based
resolutions (Ravindranath et al.\ 2001).  Extra central light results in
higher-$n$ profile fits.

Nineteen galaxies from the Balcells \& Peletier (1994, BP94) sample
were imaged with {\it HST}/NICMOS at F160W (Peletier et al.\ 1999,
Paper I).  We have performed bulge-disk decompositions, and analyze
profiles, global parameter relations and nuclear cusp slopes in a
companion paper (Balcells et al.\ 2002, Paper III).  In this Letter,
we focus on the shape indices obtained from S\'ersic fits to the bulge
profiles.  The high resolution of the {\it HST} data allows one to
disentangle nuclear disks/bars and compact sources from the extended
spheroid light.  Modeling these components, we study S\'ersic indices
for the bulge profiles and compare them to those obtained from
ground-based profiles.  We characterize the magnitude distribution of
point sources found in the galaxy centers and discuss the relevance of
low-$n$ profiles for bulge formation models.  Preliminary results on
nuclear cusps are presented in Balcells et al.\ (2001).

A key ingredient of the analysis of bulge profiles, overlooked in the
{\it HST}-based studies of bulges cited above, is the modeling of the
extended bulge and disk light distributions when fitting analytic
functions to the bulge light.  Here, composite profiles linking {\it
HST} profiles with ground-based $K$-band profiles are used so that
bulge-disk decompositions can be performed.  We assume a Hubble constant of
$H_{0} = 75$ \kms\,Mpc$^{-1}$.

\section{Data}
\label{Sec:Data}

We work with a subset of the BP94 diameter-limited sample of inclined,
early-to-intermediate type disk galaxies classified as unbarred in the
UGC (Nilson 1973); see Paper I and references therein.  None has a
Seyfert or starburst nucleus.  Ground-based surface brightness
profiles in the $K$-band for the BP94 sample, derived from ellipse
fits to UKIRT/IRCAM3 images, are given in Peletier \& Balcells (1997),
while two-dimensional bulge-disk decomposition and S\'ersic fits to
the $K$-band bulge profiles are published in APB95.

The subsample studied here comprises 19 galaxies of types S0 to Sbc
that were imaged with NICMOS on {\it HST} (camera 2,
19\arcsec$\times$19\arcsec, 0.075 arcsec/pixel).  Data reduction is
described in Paper I, while details of the data analysis are given in
Paper III. We derive elliptically-averaged surface brightness profiles
and isophotal shapes, from 0.075\arcsec to typically 10\arcsec, using
the {\tt galphot} package (J{\o}rgensen et al.\ 1992).  We keep the
centers fixed and allow the ellipticity and position angles of the
isophotes to vary.  Linking the {\it HST}-F160W and ground-based $K$
profiles is done following Paper I, with details given in Paper III.

\section{Bulge-disk fits}
\label{Sec:fits}

The combined {\it HST} plus ground-based profiles, corrected for
foreground Galactic extinction (Schlegel et al.\ 1998), $(1+z)^4$
cosmological dimming and $K$-correction, were fitted with a
PSF-convolved S\'ersic-plus-exponential model using the code from
Graham (2001) modified to use a Moffat PSF (FWHM=0.19\arcsec [NICMOS
Handbook], $\beta$=2.5).  No internal extinction corrections were
applied.  A quasi-Newton algorithm (Kahaner, Moler \& Nash 1989) was
used to converge on the solution resulting in the smallest r.m.s.\
scatter, with 5 free parameters (disk $\mu_{0}$, h; bulge $\mu_{e},
R_{e}, n$).  Each fit was individually inspected.  When an obvious
local minimum was found, the code was run again with new initial
conditions; when the apparent global minimum was found, the code was
also again run with different initial conditions to provide
confirmation.  The final fits for every galaxy are shown in
Paper III.

Such S\'ersic-exponential bulge-disk fits, covering the entire radial
range down to 0.075\arcsec, tend to show strong residuals, up to 0.3
mag, over the entire bulge.  A typical example is shown in
Figure~\ref{Fig:ExampleProfile}a.  The residual profile shows a
characteristic wave pattern.  This is caused by a strong central light
excess, as inferred from fits excluding the central 1-2 arcsec.  We
then run the fitting code including either an additional central delta
function point source (PS); a central exponential; a point source plus
an exponential; or a central Gaussian component with free FWHM to the
fitting function.  In all cases each component is PSF-convolved before
fitting.  For each galaxy we select the simplest fit, i.e. minimum
number of components, for which the residual profile falls below an
amplitude of 0.1 mag/(arcsec)$^2$.  Figure~\ref{Fig:ExampleProfile}b
shows the profile from Figure~\ref{Fig:ExampleProfile}a fitted with an
additional central PS, which yields an excellent fit.  The total light
in the inner 1-2" can also be approximated with power laws (Balcells
et al.\ 2001; B\"oker et al.  2002).  The Nuker model also fits the
inner few arcsec of the early-type galaxy profiles well (Byun et al.\
1996).  However, these approaches break down further out (e.g. Lauer
et al.\ 1995), do not allow a bulge-disk decomposition, and are not
discussed further here.  Our fits extend over 2.5 decades in radius,
and describe both bulge and disk.

\bigskip
\hskip-0.3cms\resizebox{0.9\hsize}{!}{\includegraphics{f1.eps}}
\newline{\small {\sc Fig.~\ref{Fig:ExampleProfile}.}---
(a) The combined {\it HST}+ground-based
    $H$-band surface brightness profile of NGC~5443 (Sb), with
    PSF-convolved S\'ersic bulge and
    exponential disk components (solid lines).  
    The residuals (data {\it minus} model) are shown in the lower panel. 
    (b) The same surface brightness profile fitted with a PSF-convolved 
    S\'ersic law, an
    exponential disk, plus a central point source (dashed
    line).  Residuals are shown in the lower panel.  
    
}
\bigskip
\addtocounter{figure}{1}

\begin{figure*}
    \begin{center}
    \includegraphics[height=4.5cm]{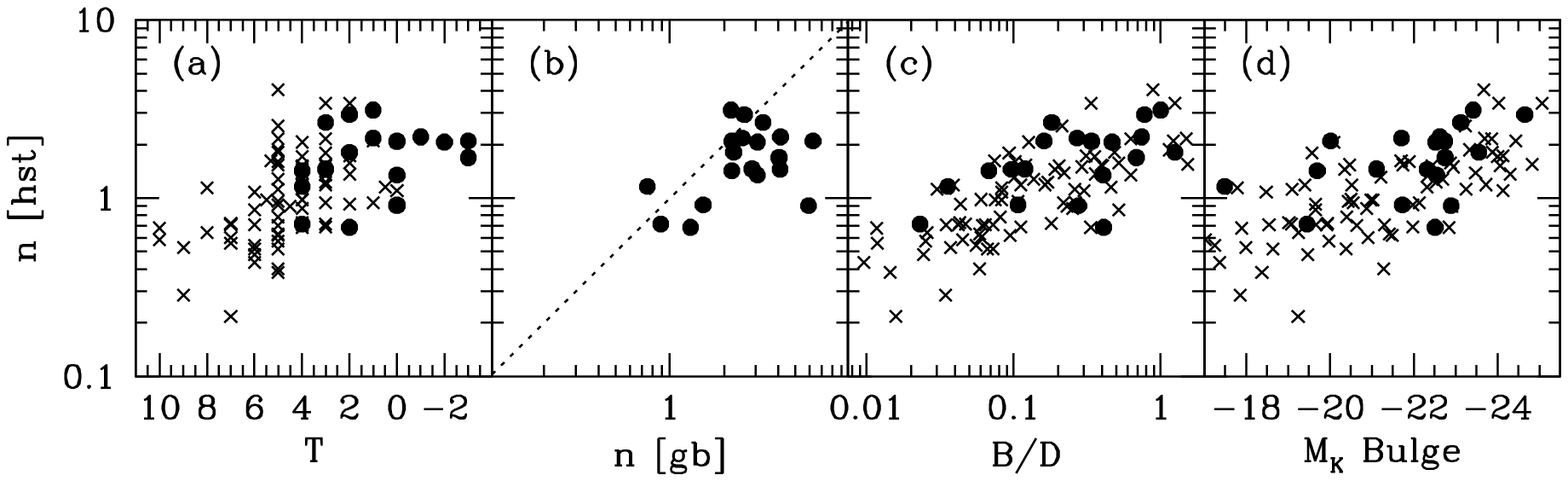}
    \end{center}
    \caption{\label{Fig:NvsTvsBDvsMKdeJ} 
    Dependence of the {\it HST}-resolved bulge S\'ersic index 
    $n_{hst}$ on global properties of the parent galaxy:
    (a) $n_{hst}$ vs galaxy morphological type index T.
    (b) $n_{hst}$ vs the bulge S\'ersic index $n_{gb}$ obtained 
        from ground-based data (APB95).
    (c) $n_{hst}$ against the bulge-to-disk ratio. 
    (d) Bulge index $n_{hst}$ against the bulge $K$-band absolute magnitude.
	{\it Filled circles: } bulges, this work. 
	{\it Crosses: } bulges from the de Jong \& van der Kruit (1994)
	sample, as analyzed by Graham (2001) (cf. his Figure~14).}
\end{figure*}

Of the 19 galaxies, 12 yield a satisfactory fit with the addition of a
single PS, 2 with the addition of a single nuclear exponential, 4 with
a PS plus an exponential, and one with a pure S\'ersic bulge.  The
frequency of PS is thus $84 \pm 8$\%, and that of inner exponentials
is $32 \pm 11$\%.  Only one galaxy (NGC~5577, Sbc) does not require an
unresolved source and can be fit with a simple S\'ersic bulge plus
exponential disk; however, the profile for this galaxy has lower S/N
than the others owing to its lower surface brightness.  Most galaxies
with PSs yield a good fit with a Gaussian inner component instead; in
those cases, the FWHM of the inner Gaussian, before PSF convolution,
is generally well below the PSF FWHM, recovering the PS solution. 
Some galaxies with inner exponential components also admit a good fit
with a marginally resolved Gaussian instead.  Because these galaxies
show high ellipticity and pointy isophotes in the nuclear region, we
choose the PS-plus-exponential fit instead of the Gaussian inner
component.  Nuclear isophotes are described in Paper III.

The S\'ersic index \nhst\ for the bulges ranges from 0.5 to 3.0, with
a mean of $<$n$> = 1.7 \pm 0.7$.  The range is significantly lower
than that obtained by APB95 using ground-based data alone, which
reaches $n=6$: Figure~\ref{Fig:NvsTvsBDvsMKdeJ}b shows that \nhst\ is
systematically lower than \ngb, especially for galaxies with
$n_{gb}>3$.  None of our bulges reaches the de Vaucouleurs \r14\
behaviour.  It appears that fits to ground-based profiles reach
S\'ersic indices $n\geq 4$ because the light from {\it HST}-unresolved
central sources, plus in some instances nuclear disks or bars, when
convolved with typical ground-based seeing, link smoothly with the the
extended bulge profile and mimic higher-$n$ S\'ersic profiles.

Figures~\ref{Fig:NvsTvsBDvsMKdeJ}c,d plot the distribution of \nhst\
vs.\ the bulge-to-disk luminosity ratio (B/D) and bulge $K$-band
absolute magnitude, respectively.  B/D are derived from the best-fit
parameters.  Bulge absolute magnitudes are derived from the galaxy
$K$-band magnitude given in APB95 and our computed B/D. As noted by
APB95, some of the galaxy magnitudes are probably slightly
underestimated when the galaxy overfills the frame. 
Figures~\ref{Fig:NvsTvsBDvsMKdeJ}c,d show that log(\nhst) correlates
with log(B/D) and with the bulge absolute $K$-band magnitude (Spearman
rank-order correlation coefficients -0.52 and -0.51, significances
97.8\% and 97.4\%, respectively).  Thus, the correlation of S\'ersic
index $n$ with B/D and bulge luminosity found in ground-based data
(APB95; Graham 2001) remains valid when fits including central PS
components are used.  We compare our {\it HST}-based $n$ distributions
to those obtained by Graham (2001) using the sample of spirals of de
Jong (1996) (Figure~\ref{Fig:NvsTvsBDvsMKdeJ}c,d,
cross symbols).  The loci occupied by the two samples are similar.  A
trend with morphological type T (Figure~\ref{Fig:NvsTvsBDvsMKdeJ}a) is
apparent when our distribution is extended to later types by adding
values from the Graham (2001) analysis.

\begin{figure*}
    \begin{center}
    \includegraphics[height=4.5cm]{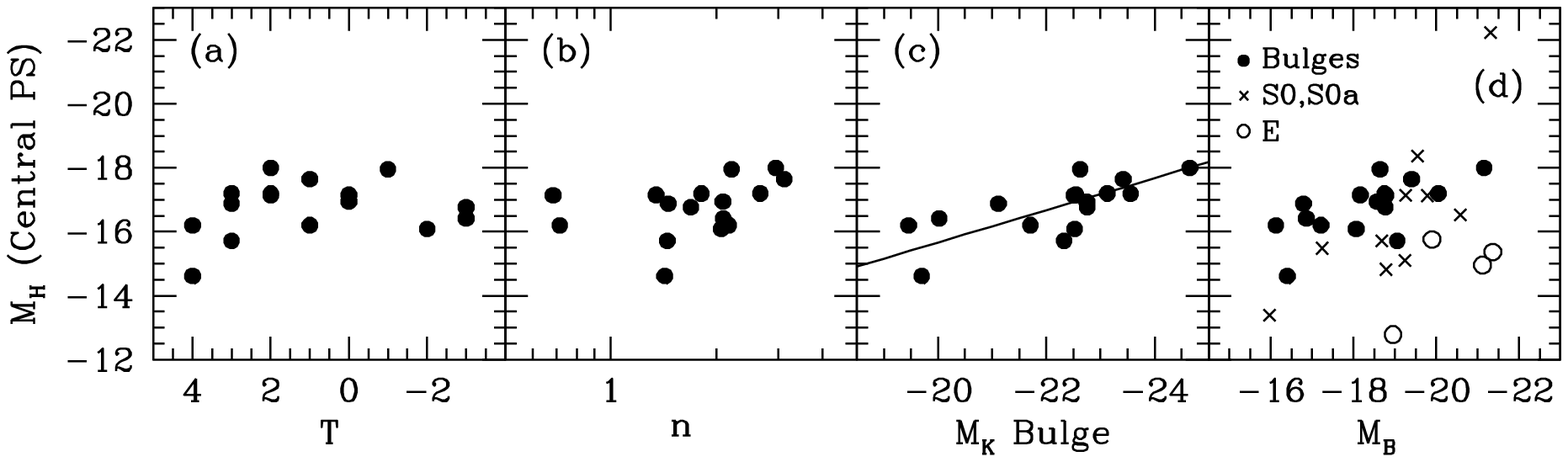}
    \end{center}
    \caption{\label{Fig:PtSrc} 
Dependence of central point source $H$-band absolute magnitude
$M_{H,PS}$ on global properties of the parent galaxy:
{\it (a)} $M_{H,PS}$ vs.\ galaxy morphological type index $T$. 
{\it(b)} $M_{H,PS}$ vs.\ S\'ersic index $n$ of the bulge. 
{\it (c)} $M_{H,PS}$ vs.\ $H$-band absolute bulge magnitude.  The line
is an orthogonal regression to the distribution.
{\it (d)} $M_{H,PS}$ vs.\ absolute $B$-band magnitude.
{\it Filled circles:} our bulges; absolute magnitude is 
that of the bulge. 
{\it Crosses:} S0 galaxies from Ravindranath et al.\ (2001).
{\it Open circles:} E galaxies from Ravindranath et al.\ (2001).
Total galaxy absolute magnitudes are used for S0 and E galaxies.
    }
\end{figure*} 

\section{Point source properties}
\label{Sec:PtSrcProp}

We showed in \S\ref{Sec:fits} that, for 16 out of the 19 galaxies
studied, adding a central PS improves the fit to the observed profile. 
Absolute $H$-band magnitudes for the PS are in the range $-14 >
M_{H,PS} > -18$.  The distribution of $M_{H,PS}$ with morphological
type T is quite flat (Figure~\ref{Fig:PtSrc}a).  $M_{H,PS}$ correlates
with $\log(n)$ and with $M_{K, Bul}$ (Spearman rank-order correlation
coefficients of -0.63 and 0.77, significances 99.6\% and 99.96\%,
respectively), see Figures~\ref{Fig:PtSrc}b,c.  The line in
Figure~\ref{Fig:PtSrc}c is an orthogonal regression to the $M_{H,PS} -
M_{K,Bul}$ relation, which yields

\begin{equation}
L_{K,PS}/L_{K,\odot} = 10^{7.36} 
(L_{K,Bul}/10^{10}\,L_{K,\odot})^{0.50\pm 0.14}.
\end{equation}

\noindent using a constant color $H-K=0.23$ for the nuclei (Persson,
Frogel \& Aaronson 1979).

Nuclear sources have been reported in previous studies of early-type
galaxy profiles based on {\it HST} imaging.  Carollo et al.\ (2002)
detect nuclear sources in 43 of their 69 late-type galaxies, with
absolute magnitudes $-10 > M_{H,PS} > -17.7$.  The bright end is
similar to ours.  The fainter luminosities are consistent with the
expected low bulge luminosity of their later type galaxies, although
the lack of a bulge-disk decomposition prevents us from comparing
theirs and our trends of PS luminosity with spheroid luminosity. 
Ravindranath et al.\ (2001), using two-dimensional fits to {\it
HST}/NICMOS F160W images of an E/S0 sample, report unresolved central
components for $\sim$50\% of their targets.  Figure~\ref{Fig:PtSrc}d
shows their distribution of point source magnitude vs.\ absolute
$B$-band galaxy magnitude together with our distribution of PS
magnitude vs.\ $B$-band absolute bulge magnitude; the latter was
computed using the $B_T$ galaxy magnitudes from the RC3 and the
bulge-disk ratio derived from the {\it HST}+ground-based profile fits. 
Their point source magnitudes are slightly fainter than ours.  Their
fainter PS magnitudes and their lower detection rate may be
consequences of their choice of the Nuker model vs.\ our use of the
S\'ersic model to fit the underlying bulge light distribution; the
Nuker law is steeper than the S\'ersic law near the origin, and
diverges at $r=0$, whereas the S\'ersic law has a finite value of
$\mu(0)$.  Also, using galaxy magnitudes for their sample instead of
bulge magnitudes shifts their points to the right in the diagram.

\section{Discussion}
\label{Sec:Discussion}

Measuring the correct value of the S\'ersic index $n$ for bulges is important
for several reasons.  Overestimating $n$ leads to an overestimate of the bulge
$r_e$ and luminosity.  Trujillo et al.\ (2001) show that fitting an $r^{1/2}$
profile with a \r14\ function yields an overestimate of $r_e$ by a factor of
3-5 and of the bulge luminosity by a factor of about 2.  The luminosity ratio
B/D is further overestimated as more of the light in the bulge-disk transition
region is assigned to the bulge.  The converse occurs when $n$ is
underestimated, eg.\ when exponential fits are forced onto $n>1$ S\'ersic
profiles (Graham 2001).  Thus, small luminosity components (typically
$L_{K,PS}/L_{K,Bul} \approx 10^{-2.8}$, eqn.\ 1) may significantly alter the
derived global parameters of bulges.  Test fits excluding inner profile regions
suggest that, in the absence of {\it HST} imaging, the effects of central
components on the bulge shape index $n$ are greatly diminished by excluding the
central FWHM of the profile, as done by Stiavelli et al.\ (2001) for dwarf
ellipticals.  Point sources should present less of a problem for late-type
spirals given the low S\'ersic indices, low bulge luminosities and hence low PS
luminosities expected from Figure~\ref{Fig:PtSrc}c.

The nuclear PSs, amounting to a fraction of order $10^{-3}$ of the bulge
luminosities, are easy to accomodate within either a merger origin or
a secular evolution origin for bulges.  The observed scaling of PS and
bulge luminosities need not indicate an internal origin for the PS as
nuclear gas deposition during a merger may scale with the masses of
the merging objects.  If bulges are evolved bars, the problem of
growing a nuclear star cluster is similar to that of feeding an active
galactic nucleus (Maciejewski et al.\ 2002).

Given the persistent description of bulges as \r14\ spheroids in galaxy
formation models, it is worth pointing out that none of the bulges follow the
\r14\ law.  APB95, de Jong (1996), Graham (2001) and MacArthur et al.\ (2002)
have shown that late- to intermediate-type bulges have $r^{1/n}$
profiles in the range $0.5 \lesssim n \lesssim 2.0$.  Our work
establishes a similar range, $0.5\lesssim n \lesssim 3.0$, for
intermediate- to early-type bulges
i.e. those which are still generally described with the \r14\ model.  
Extrapolating from the rising envelope of the $n$ -- $M_K$ relation in
Figure~\ref{Fig:NvsTvsBDvsMKdeJ}d, only bulges with $M_K \leq -26$ may reach
$n=4$.  The \r14\ law is interpreted as the consequence of violent
relaxation during a clumpy collapse (van Albada 1982), and is obtained after
mergers of similar-size disk galaxies (Barnes 1988).  Simulations of
collisionless satellite accretion onto bulge-disk galaxies further show that
bulges evolve smoothly from exponential profiles to \r14\ profiles when
doubling their mass due to accretion events (ABP01).  At face value then, our
distribution of S\'ersic shape indices $n$
(Figure~\ref{Fig:NvsTvsBDvsMKdeJ}c,d) suggests that the vast majority of bulges
did not acquire their present structural properties by growing through
collisionless mergers.  Kinematic data might show whether, besides a low
S\'ersic index $n$, these bulges show other properties typical of disks
(Kormendy 1993; Falc\'on-Barroso et al.\  2002).

%The effect of gas
%dynamics and ensuing star formation in the mergers would presumably be
%to boost the central density (Barnes \& Hernquist 1991), yielding a
%higher-$n$ value than obtained from collisionless dynamics alone,
%hence deviating even more from the observed range of shape indices
%$n$.

Have deviations from the \r14\ law in $N$-body models gone
undetected?  S\'ersic fitting is only starting to be applied in
$N$-body studies.  Hjorth \& Madsen (1995) show that violent
relaxation in a finite volume, combined with subsequent escape of
positive-energy stars, leads to a distribution function with
deviations from the \r14\ law; from their figures, a shallow central
potential might possibly fit $n<4$ S\'ersic profiles.  
%This situation
%could apply to the merger growth of small bulges embedded in massive
%disks, a configuration that ABP01 do not model.  
Accretion of low-mass
satellites might heat the bulge only moderately and deposite the
compact satellite nucleus at the center of the remnant.  This could
lead to a $n\sim 1-2$ bulge with a central brightness spike not unlike
what is observed.  But very long merger times and slow growth rate for
low-mass satellites are a problem for this hypothesis.  Recent
cosmological SPH simulations of galaxy formation (Scannapieco \&
Tissera 2002) find $n\approx 1$ for bulges resulting from secular
evolution, while merging rises $n$ to about $n=4$, again pointing at 
low-$n$ profiles being related to non-merger formation histories.

\end{document}